\documentclass[twoside]{article}
\usepackage[T1]{fontenc}
\usepackage{graphicx}
\usepackage{amsmath}
\newcommand{\J}{\mathcal{J}}
\renewcommand{\vec}[1]{\boldsymbol{#1}}

\catcode`\@=11
\long\def\@makefntext#1{
\protect\noindent \hbox to 3.2pt {\hskip-.9pt
$^{{\eightrm\@thefnmark}}$\hfil}#1\hfill}		

\def\@makefnmark{\hbox to 0pt{$^{\@thefnmark}$\hss}}	

\def\ps@myheadings{\let\@mkboth\@gobbletwo
\def\@oddhead{\hbox{}
\rightmark\hfil\eightrm\thepage}
\def\@oddfoot{}\def\@evenhead{\eightrm\thepage\hfil
\leftmark\hbox{}}\def\@evenfoot{}
\def\sectionmark##1{}\def\subsectionmark##1{}}

\oddsidemargin=\evensidemargin
\newcounter{sectionc}\newcounter{subsectionc}\newcounter{subsubsectionc}
\renewcommand{\section}[1] {\vspace{12pt}\addtocounter{sectionc}{1}
\setcounter{subsectionc}{0}\setcounter{subsubsectionc}{0}\noindent
	{\tenbf\thesectionc. #1}\par\vspace{5pt}}
\renewcommand{\subsection}[1] {\vspace{12pt}\addtocounter{subsectionc}{1}
	\setcounter{subsubsectionc}{0}\noindent
	{\bf\thesectionc.\thesubsectionc. {\kern1pt \bfit #1}}\par\vspace{5pt}}
\renewcommand{\subsubsection}[1] {\vspace{12pt}\addtocounter{subsubsectionc}{1}
	\noindent{\tenrm\thesectionc.\thesubsectionc.\thesubsubsectionc.
	{\kern1pt \tenit #1}}\par\vspace{5pt}}
\newcommand{\nonumsection}[1] {\vspace{12pt}\noindent{\tenbf #1}
	\par\vspace{5pt}}

\newcounter{appendixc}
\newcounter{subappendixc}[appendixc]
\newcounter{subsubappendixc}[subappendixc]
\renewcommand{\thesubappendixc}{\Alph{appendixc}.\arabic{subappendixc}}
\renewcommand{\thesubsubappendixc}
	{\Alph{appendixc}.\arabic{subappendixc}.\arabic{subsubappendixc}}

\renewcommand{\appendix}[1] {\vspace{12pt}
        \refstepcounter{appendixc}
        \setcounter{figure}{0}
        \setcounter{table}{0}
        \setcounter{lemma}{0}
        \setcounter{theorem}{0}
        \setcounter{corollary}{0}
        \setcounter{definition}{0}
        \setcounter{equation}{0}
        \renewcommand{\thefigure}{\Alph{appendixc}.\arabic{figure}}
        \renewcommand{\thetable}{\Alph{appendixc}.\arabic{table}}
        \renewcommand{\theappendixc}{\Alph{appendixc}}
        \renewcommand{\thelemma}{\Alph{appendixc}.\arabic{lemma}}
        \renewcommand{\thetheorem}{\Alph{appendixc}.\arabic{theorem}}
        \renewcommand{\thedefinition}{\Alph{appendixc}.\arabic{definition}}
        \renewcommand{\thecorollary}{\Alph{appendixc}.\arabic{corollary}}
        \renewcommand{\theequation}{\Alph{appendixc}.\arabic{equation}}
        \noindent{\tenbf Appendix \theappendixc #1}\par\vspace{5pt}}
\newcommand{\subappendix}[1] {\vspace{12pt}
        \refstepcounter{subappendixc}
        \noindent{\bf Appendix \thesubappendixc. {\kern1pt \bfit #1}}
	\par\vspace{5pt}}
\newcommand{\subsubappendix}[1] {\vspace{12pt}
        \refstepcounter{subsubappendixc}
        \noindent{\rm Appendix \thesubsubappendixc. {\kern1pt \tenit #1}}
	\par\vspace{5pt}}

\newcommand{\textlineskip}{\baselineskip=13pt}
\newcommand{\smalllineskip}{\baselineskip=10pt}

\def\eightcirc{
\begin{picture}(0,0)
\put(4.4,1.8){\circle{6.5}}
\end{picture}}
\def\eightcopyright{\eightcirc\kern2.7pt\hbox{\eightrm c}}

\newcommand{\copyrightheading}[1]
	{\vspace*{-2.5cm}\smalllineskip{\flushleft
	{\footnotesize International Journal of Modern Physics C, #1}\\
	{\footnotesize $\eightcopyright$\, World Scientific Publishing
	 Company}\\
	 }}

\newcommand{\publisher}[2]{{\begin{center}\footnotesize\smalllineskip
	Received #1\\
	Revised #2
	\end{center}
	}}

\def\abstracts#1#2#3{{
	\centering{\begin{minipage}{4.5in}\baselineskip=10pt\footnotesize
	\parindent=0pt #1\par
	\parindent=15pt #2\par
	\parindent=15pt #3
	\end{minipage}}\par}}

\def\keywords#1{{
	\centering{\begin{minipage}{4.5in}\baselineskip=10pt\footnotesize
	{\footnotesize\it Keywords}\/: #1
	\end{minipage}}\par}}

\renewenvironment{thebibliography}[1]
        {\frenchspacing
	 \ninerm\baselineskip=11pt
         \begin{list}{\arabic{enumi}.}
        {\usecounter{enumi}\setlength{\parsep}{0pt}
	 \setlength{\leftmargin 12.7pt}{\rightmargin 0pt} 
         \setlength{\itemsep}{0pt} \settowidth
	{\labelwidth}{#1.}\sloppy}}{\end{list}}

\newcounter{itemlistc}
\newcounter{romanlistc}
\newcounter{alphlistc}
\newcounter{arabiclistc}

\newcommand{\fcaption}[1]{
        \refstepcounter{figure}
        \setbox\@tempboxa = \hbox{\footnotesize Fig.~\thefigure. #1}
        \ifdim \wd\@tempboxa > 5in
           {\begin{center}
        \parbox{5in}{\footnotesize\smalllineskip Fig.~\thefigure. #1}
            \end{center}}
        \else
             {\begin{center}
             {\footnotesize Fig.~\thefigure. #1}
              \end{center}}
        \fi}

\newcommand{\tcaption}[1]{
        \refstepcounter{table}
        \setbox\@tempboxa = \hbox{\footnotesize Table~\thetable. #1}
        \ifdim \wd\@tempboxa > 5in
           {\begin{center}
        \parbox{5in}{\footnotesize\smalllineskip Table~\thetable. #1}
            \end{center}}
        \else
             {\begin{center}
             {\footnotesize Table~\thetable. #1}
              \end{center}}
        \fi}

\def\@citex[#1]#2{\if@filesw\immediate\write\@auxout
	{\string\citation{#2}}\fi
\def\@citea{}\@cite{\@for\@citeb:=#2\do
	{\@citea\def\@citea{,}\@ifundefined
	{b@\@citeb}{{\bf ?}\@warning
	{Citation `\@citeb' on page \thepage \space undefined}}
	{\csname b@\@citeb\endcsname}}}{#1}}

\newif\if@cghi
\def\cite{\@cghitrue\@ifnextchar [{\@tempswatrue
	\@citex}{\@tempswafalse\@citex[]}}
\def\citelow{\@cghifalse\@ifnextchar [{\@tempswatrue
	\@citex}{\@tempswafalse\@citex[]}}
\def\@cite#1#2{{$\null^{#1}$\if@tempswa\typeout
	{IJCGA warning: optional citation argument
	ignored: `#2'} \fi}}

\def\pmb#1{\setbox0=\hbox{#1}
	\kern-.025em\copy0\kern-\wd0
	\kern.05em\copy0\kern-\wd0
	\kern-.025em\raise.0433em\box0}

\def\fnt#1#2{\footnotetext{\kern-.3em
	{$^{\mbox{\scriptsize #1}}$}{#2}}}

\def\fpage#1{\begingroup
\voffset=.3in
\thispagestyle{empty}\begin{table}[b]\centerline{\footnotesize #1}
	\end{table}\endgroup}

\def\runninghead#1#2{\pagestyle{myheadings}
\markboth{{\protect\footnotesize\it{\quad #1}}\hfill}
{\hfill{\protect\footnotesize\it{#2\quad}}}}
\font\tenrm=cmr10
\font\tenit=cmti10
\font\tenbf=cmbx10
\font\bfit=cmbxti10 at 10pt
\font\ninerm=cmr9

\font\eightrm=cmr8

\input{psfig.sty}
\usepackage{amsmath}

\begin{document}

\runninghead{F.~Bagnoli \& M.~Bezzi}{Species formation in simple ecosystems}
\normalsize\textlineskip
\thispagestyle{empty}
\setcounter{page}{1}
\copyrightheading {Vol. 0, No. 0 (1998) 000--000}
\vspace*{0.88truein}
\fpage{1}

\centerline{\bf SPECIES FORMATION IN SIMPLE ECOSYSTEMS}
\vspace*{0.37truein}
\centerline{\footnotesize F.~BAGNOLI}
\vspace*{0.015truein}
\centerline{\footnotesize\it Dipartimento di Matematica Applicata}
\centerline{\footnotesize\it Universit\`a di Firenze}
\centerline{\footnotesize\it via S. Marta 3, I-50139, Firenze, Italy}
\centerline{\footnotesize\it INFN and INFM, Sez. di Firenze}
\centerline{\footnotesize\it E-mail: bagnoli@dma.unifi.it}
\vspace*{0.15truein}
\centerline{\footnotesize and}
\vspace*{0.15truein}
\centerline{\footnotesize M.~BEZZI}
\vspace*{0.015truein}
\centerline{\footnotesize\it Dipartimento di Fisica}
\centerline{\footnotesize\it Universit\`a di Bologna}
\centerline{\footnotesize\it via Irnerio 46, I-40126 Bologna, Italy}
\centerline{\footnotesize\it INFN, sez. Bologna}
\centerline{\footnotesize\it E-mail: bezzi@ing.unifi.it}

\vspace*{0.225truein}
\publisher{(received date)}{(revised date)}
\vspace*{0.21truein}

\abstracts{
In this paper we consider a microscopic model of a simple ecosystem.
The basic ingredients of this model  are individuals,
and both the phenotypic and genotypic levels are taken in account.  
The model is based on a long range cellular automaton (CA);
introducing simple interactions between the individuals,
we get some of the complex collective behaviors  
observed in a real ecosystem.
Since our fitness function is smooth, the model does not exhibit the error
threshold transition; on the other hand
the size of total population is not kept constant, and the mutational
meltdown transition is present. 
We study the effects of competition between genetically similar individuals
and how it can lead to species formation. This speciation transition
does not depend on the mutation rate. We present also an analytical
approximation of the model. 
}{}{}
\vspace*{10pt}
\vspace*{1pt}\textlineskip

\section{Introduction\label{sec:Intro}}

Real ecosystems present a complex behavior. Many of their peculiar
features are considered in classical population dynamics models, where the
dynamical variables are the number of individuals of different
populations.\cite{Murray}
In this paper a different point of view is considered, 
we propose a microscopic model of an evolving (in Darwinian sense)
 ecosystem, where the individuals are represented by their genotypes.
Our model is related to the Eigen model
for quasispecies,\cite{Eigen71,Eigen:quasispecies,Peliti}
although we consider a different fitness landscape and
the presence of interactions among individuals. 
  
Considering simple ecological interactions (competition, predation, 
cooperation), we are able to obtain
a complex collective behavior. The aim of this work is to have a simple
predictive model, that can reproduce some of the basic features present in an
real ecosystem, such as:  
\begin{itemize}
\item Evolution. The system has to be able to
create diversity (mutations). Darwinian selection acts on this 
diversity.
\item Population dynamics. The model should reproduce the typical
phenomenology of population dynamics, such as logistic growth for single species
dynamics in limited environment, Lotka-Volterra dynamics for predator-prey
interactions, etc.
\item Self organized behavior. One expect to observe collective phenomena such
as trophic chain formation or species formation.     
\item Response to external stimuli, in particular to environmental 
changes such as cycle of seasons or human intervention.   
\end{itemize}

In classical population dynamics the building block are the species,
and the interactions among them. Since
we want to study the self-organization of an ecosystem (including
species formation) we take as a building block the single individual.

In our schematization the individual is identified by its genotype
$x$, which is represented as a fixed length string of $L$ bits:
the genotype space is a Boolean hypercube of $L$ dimensions,
and mutations correspond to displacements in this space.
On the other hand, a genotype identifies a strain of individuals.

Individuals are able to survive according with a fitness function, which
also takes into account the interactions with other individuals  in the
environment.
Natural selection however does not act directly on the genotype, 
but rather 
on the resulting phenotype $g(x)$, which can
be considered a function of the genotype $x$.\footnote{The assumption that 
the phenotype is a single-valued function
of the genotype implies that we are not
considering polymorphism (the fact that two cells with the same
genotype can have different morphologies) nor age structure.}
 Generally, the
phenotypic space is simpler than the genotypic one, according with
the number of morphological characters considered. 
In our simple ecosystem model, $g(x)$
is simply the fraction of ones in $x$.\footnote{One can assume that 
in each locus there are two alleles of a given gene: 0 for the ``good''
allele and 1 for the bad one}~ In this case the
phenotypic space is one-dimensional.  The smoothness of the fitness
function is related to the resolution required in genotypic space: if
one clusters together a species into a single point then the fitness
function can be quite rough. Since we are interested in the phenomenon of
species formation, we require a smooth fitness landscape.  

Finally, there is the real space.  
We shall describe in Section~2 a one-dimensional 
cellular automaton (CA) model. However, we shall consider 
only  the limit of very long interaction
length, i.e.\ global coupling. This simplification allows us to separate
 the complexity of the dynamics in
genotypic space  from spatial pattern formation.

Similar systems have been introduced
in order to investigate
the phenomenon for which a
phenotypically favored strain can loose its predominance due to a high
mutation rate (error threshold).~\cite{Eigen71,Eigen:quasispecies,Peliti}~
In these works the population size is kept constant;
recent preliminary results~\cite{Malarz} shows that if the population
size is allowed to fluctuate (limited by an external constraint)
another transition, called mutational
meltdown~\cite{MutationalMeltdown,Bernardes}
can be observed.  In this case the whole population vanishes while not
changing its distribution. No direct competition
among individuals is considered.

We are mainly interested in the problem of species formation due to
inter-individual competition, in the limit of very small mutation rates. 
For this reason (and also due to the smoothness of the static fitness
function),
in our model the error threshold transition is not observable. 
Moreover, we do not impose any limit on the population size:
individuals compete for free space and this automatically limits the
size of population. 
The free space limitation translates into a logistic-like
equation for the whole population size, and this furnishes a simple
illustration of the mutational meltdown transition (see Section~3).

Analytic approximations can be obtained if one takes into consideration only
the simpler phenotypic space, as reported in Appendix~A. We are able
to compute the speciation threshold, for which the population
distribution splits into several separated peaks also for a very
smooth fitness landscape and the mutational meltdown threshold.

We developed an optimized computer algorithm for the original model, 
see Appendix~B. The results of the
simulations are reported in Section~3; the speciation transition
appears, for a choice of parameters consistently with our analytical
results. In this more ``realistic'' case, the population 
distribution is not at all trivial, exhibiting coexistence of several quasi
species at the same distance from the fittest strain.      

\section{The cellular automaton model\label{sec:Model}}

Let us consider an early ecosystem, 
populated by haploid\footnote{Single copy of 
genetic material, thus non sexually reproducing.}~ individuals. 
Each individual occupies 
a cell of a lattice in an one dimensional space;
the size of the lattice is $N$ sites. Each individual
 is identified
by its genetic information (genotype),
 that we model as a base two number $x$ of  $L$ bits.
The  distance in the genotypic space is defined in
terms of the number of mutations needed to connect (along the shortest
path) two individuals. We shall consider only point mutations
($0\leftrightarrow 1$), occurring  with
probability $\mu$ independently of the bit position. Thus the
genotypic distance $d(x,y)$ between strains $x$ and $y$ 
is simply their Hamming distance (number of different bits). 
The mutation probability $W(x,y)$ is
\[
	W(x,y) = \mu^{d(x,y)}(1-\mu)^{L-d(x,y)},
\] 
which for vanishing mutation  rate $\mu \rightarrow 0$ can be written
in a quasi-diagonal form
\[
\begin{array} {rcll}
 W(x,y) &= & \mu\qquad& \mbox{if $d(x,y)=1$} \\
 W(x,x) &=& 1-L\mu & \\
 W(x,y) &=&0&\mbox{otherwise.}
\end{array}
\]
 
Given a genotype $x$, its phenotype is represented by a function
$g(x)$, which maps the genotypic space into the phenotypic one. 
In this paper we shall consider 
a very simple mapping, $g(x)= d(x,0)$, i.e.\ the phenotype is
proportional to the number of ones in the genotype.
 
This automaton has a large number of states, one for each different genome
plus a state ($*$) representing the empty cell. 
The evolution of the system is given by the
application of two rules: the survival step, that includes 
the interactions among individuals,
and the reproduction step.  

{\bf Survival}: An individual $x_i\equiv x_i^t \neq *$  
at time $t$ and site $i$, $i=1,\dots,N$,  
has a probability $\pi$ 
of surviving per unit of time. It is reasonable to assume this probability 
to depend only on phenotypic characters. 
The survival probability $\pi=\pi(H)$ is expressed as a sigma-shaped function
of  the fitness function $H$:\footnote{Our choice of the fitness function
does not consider the reproductive efficiency.}
\begin{equation}
\label{probability}
  \pi(H) = \dfrac{e^{\beta H}}{1+e^{\beta H}}=
     \dfrac{1}{2} + \dfrac{1}{2} \tanh(\beta H),
\end{equation}
where $\beta$ is a parameter that can be useful to modulate the effectiveness
of selection. We always use $\beta=1$.  
We define the fitness $H$
of the strain $x_i$ in the environment $\vec{x}=\{x^t_{1}, \dots,
x^t_{N}\}$ as
 \begin{equation}
 \label{fitnessH}
  H(x_i,\vec{x}) = h(g(x_i)) + 
   \dfrac{1}{N}\sum_{j=1}^{N} \J(g(x_i),g(x_j)).
\end{equation}

The fitness function is composed by two parts: 
the static fitness $h(g(x_i))$, 
and the interaction term
$1/N\sum_{j=1}^{N} \J(g(x_i),g(x_j))$.
The matrix
$\J$  define the chemistry of the world and is fixed; 
the field $h$ represents the fixed or slowly changing environment.  
A strain $x$ with static fitness
$h(g(x))>0$ represents individuals that can survive in isolation
(say, after an inoculation into an empty substrate),  while a strain with
$h(g(x))<0$ represents predators or parasites that requires the presence
of some other individuals to survive. The interaction matrix $\J$ specifies the
inputs for non autonomous strains.

We assume that the static fitness $h(u)$ is a 
linear decreasing function of $u$
except in the vicinity of $u=0$, where it has a quadratic maximum:
\begin{equation}
	h(u) = h_0 + b\left(
		1-\dfrac{u}{r} - \dfrac{1}{1+u/r}
	\right)\label{H0}
\end{equation}
so that close to $u=0$ one has 
$h(u) \simeq h_0-b u^2/r^2$ and for $u\rightarrow \infty$,
$h(u) \simeq h_0+b(1-u/r)$. 
Thus, the master sequence (in Eigen's language) is located at $x=0$.

The matrix $\J$ mediates the interactions between two strains. 
For a classification in terms of usual
ecological interrelations, one has to consider together $\J(u,v)$ and
$\J(v,u)$.
One can have four cases:
\vspace{.3cm}
\begin{center}
\begin{tabular}{ccl}
$\J(u,v)<0$ & $\J(v,u)<0$ & competition \\
$\J(u,v)>0$ & $\J(v,u)<0$ & predation or parasitism \\
$\J(u,v)<0$ & $\J(v,u)>0$ & predation or parasitism \\
$\J(u,v)>0$ & $\J(v,u)>0$ & cooperation 
\end{tabular}
\end{center}
\vspace{.3cm}
  
Since the individuals with similar phenotypes are those sharing the largest
quantity of resources, the competition is stronger the more similar their
phenotypes are (intraspecies competition). 
This implies that the interaction matrix $\J$ has negative components
near the diagonal. We do not include here neither
familiar structures nor sexual mating between
genetically akin individuals nor other kind of competition or cooperation. 

We have chosen following form for the interaction matrix $\J$:
\begin{equation}
  \J(u,v) = - J  K\left(
			\dfrac{u-v}{R}\right)  
\label{J}
\end{equation}
with the kernel $K$ given by
\[
	K(r) = \exp\left(-\dfrac{|r|^\alpha}{\alpha}\right),
\]
i.e. a symmetric decreasing function of $r$ with 
$K(0)=1$. The parameter $J$  and $\alpha$ control
the intensity and the steepness of the intraspecies competition,
respectively. We shall use a Gaussian ($\alpha=2$) kernel, for the
motivations illustrated in Appendix~A.    

The survival phase is thus expressed as:
\begin{itemize}
\item If $x_i \ne *$ then we get, with probability $\pi(H(x_i, \vec{x}))$,
$x'_i=x_i$,  otherwise $x'_i= * $ 
\item Else	$x'_i = x_i = *$
\end{itemize}

{\bf Reproduction}: The reproductive phase can be implemented as a rule for
empty cells: choose randomly one of the neighboring cells 
 and copy its state; if it is different from the empty state then 
 apply mutations by
reversing the value of one bit with probability $\mu$. 

One can notice that 
the effective reproduction rate does not only depend  on the
survival probability of the individual,
but also on total availability of empty cells.

One can have an insight of the features of the model by a simple mean
field analysis. 
Let $n(x)$ be  the number of organisms with genetic code $x$, and $n_*$ the
number of empty sites, 
\[
	n_*+\sum_xn(x) =N.
\]
We denote with  $m$ 
the relative abundance of non-empty sites: 
\[
m ={\sum_xn(x)}/N =1-n_*/N.
\] 
The sums do not run over the  empty cell state ($x\neq *$).
We can express the fitness function $H$ (and thus the survival
probability $\pi$) in terms of the number of individuals $n(x)$ 
in a given strain or in terms of the probability distribution $p(x)=n(x)/mN$
\begin{eqnarray}
	H(x,\vec{n}) &=& h(x)+\frac 1N\sum_yJ(x,y)n(y);\label{Hn}\\
	H(x,\vec{p})&=& h(x)+m\sum_yJ(x,y)p(y).\label{Hp}
\end{eqnarray}

The average evolution of the system  will be governed
by the following equations, in which a tilde labels quantities after
the survival step, and a prime after the reproduction step:
\begin{eqnarray}
	\tilde n(x) &=&\pi(x,\vec{n})n(x),  \label{ntilde} \\
	n^{\prime }(x) &=&\tilde n(x)+\frac{\tilde n_*}N\sum_yW(x,y)\tilde n(y). 
		\label{nprime}
\end{eqnarray}

Using the properties of $W$,
\[
	\sum_yW(x,y)=\sum_xW(x,y)=1,
\]
and summing over $x$ in Eqs.~(\ref{ntilde}) and (\ref{nprime}),
 we obtain an equation for $m$:
\begin{eqnarray}
	\widetilde{m} &=&\frac{\sum_x\tilde n(x)}N=
	 \frac 1N\sum_x\pi(x,\vec{n})n(x)=
	 m\overline{\pi} \label{mtilde}\\
	m^{\prime } &=&\frac{\sum_xn^{\prime }(x)}N=
	 \widetilde{m}+\frac{\tilde n_*}{%
N^2}\sum_{y}\tilde n(y), \label{mprime}
\end{eqnarray}
i.e.
\begin{equation}
 	m'=m\overline{\pi}(2-m\overline{\pi}), \label{logistic}
\end{equation}
where
\[
	\overline{\pi}\equiv \frac
    1{mN}\sum_x\pi(x,\vec{n})n(x)=\sum_x\pi(x,\vec{p})p(x)
\]
is the average survival probability.

The normalized evolution equation for $p(x)$ is:
\begin{equation}
	p^{\prime }(x)=\frac{
	\pi(x,p,m)p(x)+(1-m\overline{\pi})\sum_yW(x,y)\pi(y,p,m)p(y)}
	{\overline{\pi}(2-m\overline{\pi})}.   \label{peq}
\end{equation}

Notice that Eq.~({\ref{logistic}}) is a logistic equation with $\overline{\pi}$ 
as control parameter.  The stationary condition, ($m^{\prime }=m$), is
\begin{equation}
	m =\frac{2\overline{\pi}-1}{\overline{\pi}^2}.
	\label{mequation}
\end{equation}

One observes extinction if $\overline{\pi} \le 1/2$. 
The decrease of $\overline{\pi}$ can arise from a variation of the
environment (notably $h(x)$) or from an increase of the mutation rate
$\mu$, which broadens the distribution $p(x)$.
This last effect 
corresponds to
the mutational meltdown, for which the population vanishes while
continuing to exhibits a quasi-species distribution. 
Since the total
population $m$ multiplies the competition term in Eq.~(\ref{peq}), one
cannot observe coexistence of species due to competition near the
mutational meltdown transition. 
From Eq.~({\ref{logistic}})
one could expect a periodic
or chaotic behavior of the population; 
however, since $\overline{\pi}$ is always less than one, the asymptotic
dynamics of the 
population $m$ can only exhibit fixed points. 

We are mainly interested in the asymptotic behavior of the population 
in the limit
$\mu\rightarrow 0$. Actually, the mutation mechanism is needed only to
define
the genetic distance and to populate an eventual niche.
The results should not change qualitatively if  more realistic
mutation mechanisms are included.

Let us first examine the behavior of Eq.~(\ref{peq}) in absence of
competition ($J=0$) for a smooth static landscape and  a vanishing
mutation rate. This corresponds to the Eigen
model,\cite{Eigen71,Eigen:quasispecies} with a
smooth fitness landscape.  Since it does not exhibit any phase
transition, the asymptotic distribution is unique.  The asymptotic
distribution is given by one delta function peaked around the global
maximum of the static landscape, or more delta functions 
(coexistence) if the global maxima are degenerate.  The effect of a
small  but finite mutation rate is simply that of broadening the
distribution from a delta peak to a bell-shaped
curve~\cite{Bagnoli:ecal} (quasi-species).

\begin{figure}[t]
\centerline{\psfig{figure=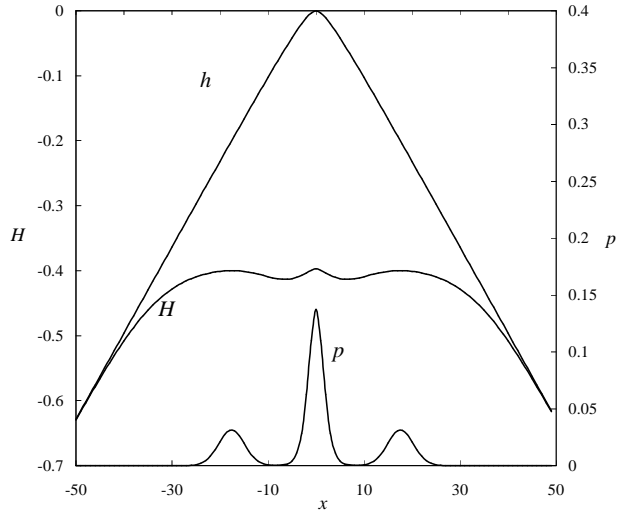,width=8cm}}
\caption{\label{fitness} 
Static fitness $h$, effective fitness 
$H$,  and asymptotic distribution $p$ 
for the phenotypic model studied in Bagnoli and Bezzi
(1997)~\protect\cite{BagnoliBezzi}~,analogous to
Eq.~(\protect\ref{peq}).}
\end{figure}

While the degeneracy of maxima of the  static fitness
landscape is a very particular condition, 
we shall show in Appendix A that in
presence of competition this is a generic case. 

For illustration, we report 
in Figure~\ref{fitness} 
the asymptotic distribution of the population and the static and
effective fitness
for a similar model~\cite{BagnoliBezzi}
in which the genotypic space is approximated by a phenotypic one
(see Appendix A for details).
The effective fitness $H$ is here almost 
degenerate, since  $\mu$ is greater than zero 
and the competition effect extends on
the neighborhood of
the maxima, and this leads to the coexistence.

\begin{figure}[t]
\centerline{\psfig{figure=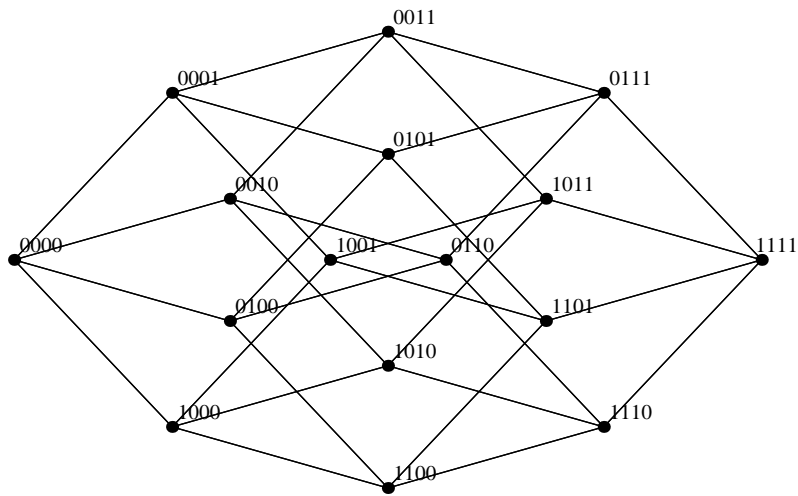,width=12cm,angle=270}}
\caption{\label{fig:Hyper} The representation of the Boolean hypercube for
$L=4$}
\end{figure}

\begin{figure}[t]
\centerline{\psfig{figure=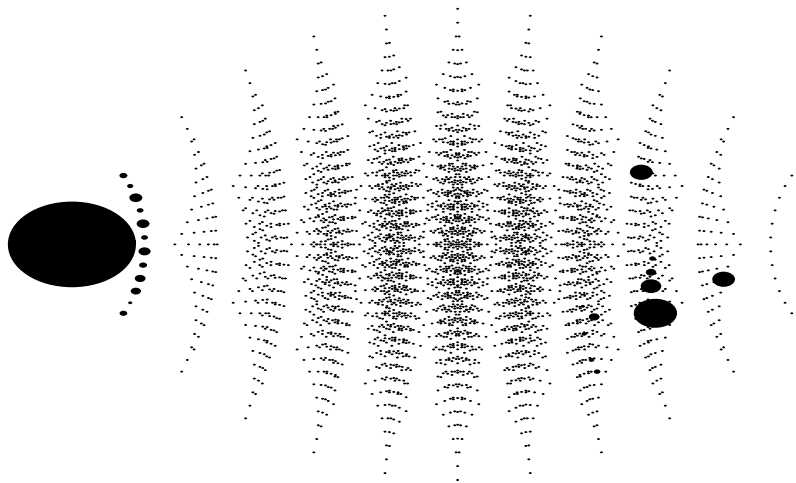,width=12cm,angle=270}}
\caption{\label{fig:Spot} Quasispecies in hypercubic space for
$L=12$. The smallest points represent placeholder of strains (whose
population is less than $2\cdot 10^-2$), only the
larger dots corresponds to effectively populated quasispecies; the size of the
dot is proportional to the square root of population. Parameters:
$\mu=10^{-3}$, $h_0=2$, $b=10^{-2}$, $R=5$, $r=0.5$, $J=0.28$,
$N=10000$, $L=12$.}
\end{figure}

\begin{figure}[h]
\centerline{\psfig{figure=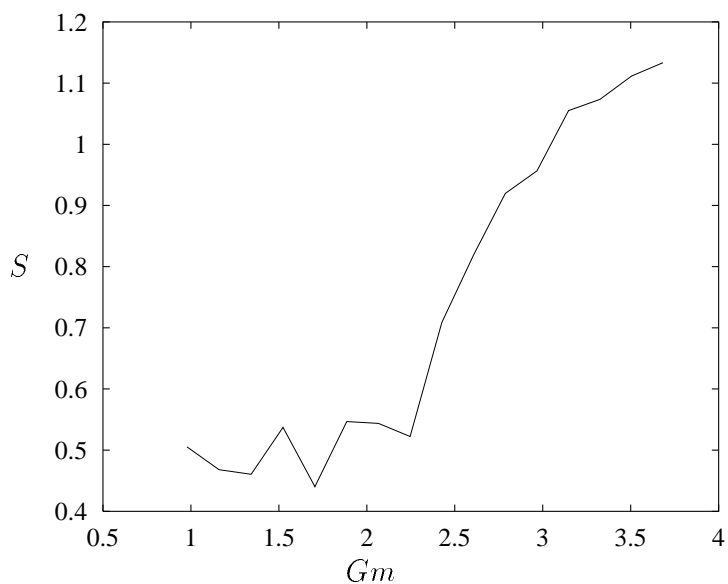,width=10cm}}
\caption{\label{fig:Entropy} The speciation transition characterized
by the entropy $S$ as a function of the control
parameter $G\;m$. Each point is an average over 15 runs.  Same parameters
as in Figure~\protect\ref{fig:Spot}, varying $J$.}
\end{figure}

\begin{figure}[h]
\centerline{\psfig{figure=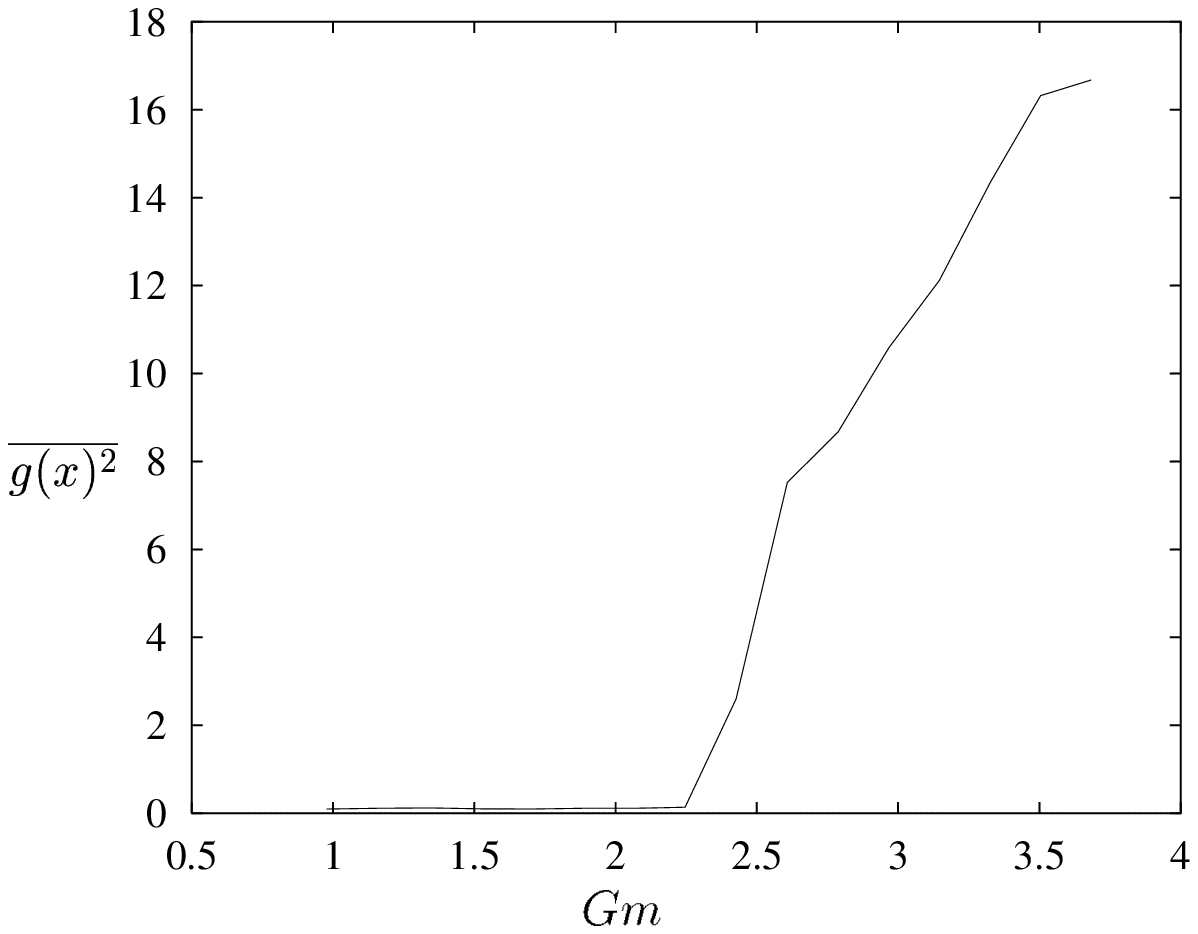,width=6.5cm}
\psfig{figure=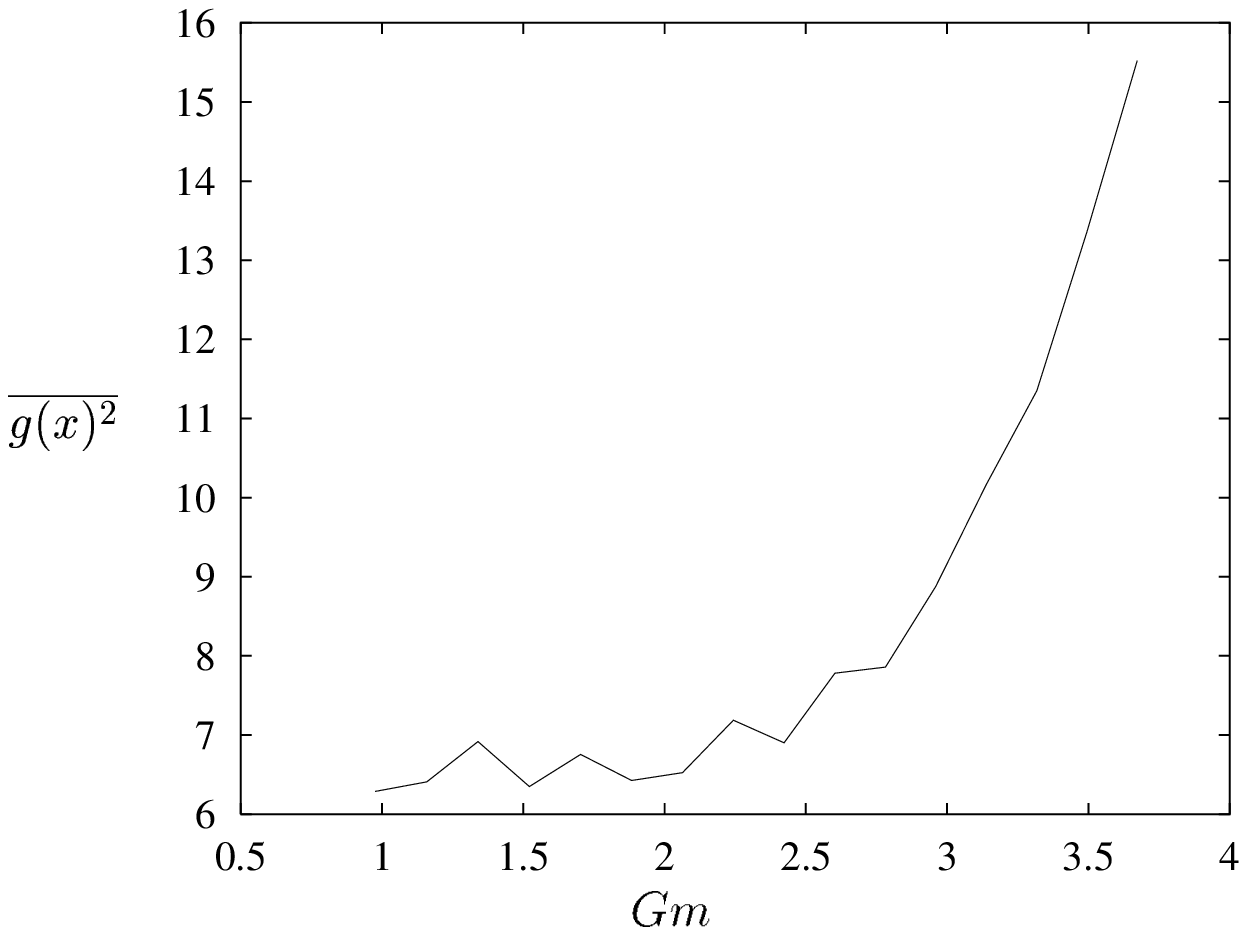,width=6.5cm}}
\caption{\label{fig:x2} Independence of 
the speciation transition by the mutation rate. 
The transition is characterized by
the average square phenotypic distance $\overline{g(x)^2}$ 
of distribution $p(x)$, as a function of the control
parameter $G\;m$. Each point is a single run.  
Same parameters
as in Figure~\protect\ref{fig:Spot}, varying $J$ with $\mu=10^{-3}$
(left) and $\mu=5\;10^{-2}$ (right). }
\end{figure}

\begin{figure}[t]
\centerline{\psfig{figure=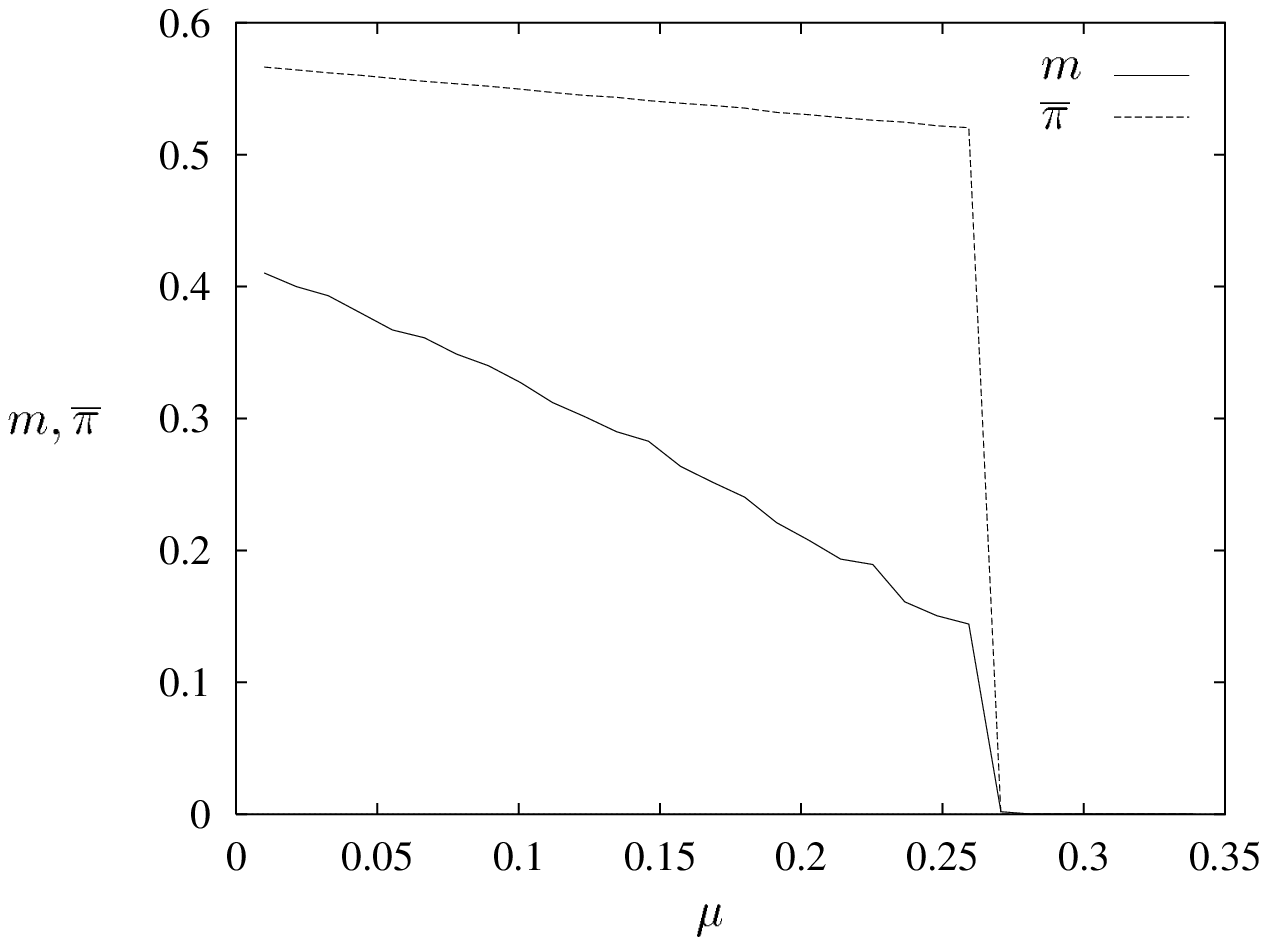,width=10cm}}
\caption{\label{fig:melt} Meltdown transition  characterized by
$m$  and $\overline{\pi}$ as a function of $\mu$. Here $J=0.3$,
$h_0=0.4$, $b=0.35$, $r=0.5$, $R=5$, $N=2000$, $L=8$.}
\end{figure}

\section{Speciation and mutational meltdown in hypercubic genotypic space}
We have developed an optimized code (reported in Appendix~B) for the 
simulation of the original model.
We  use the following \emph{easter egg}\footnote{We acknowledge D. Stauffer
for having suggested this name.}~ representation for 
quasi-species  in hypercubic space: 
starting from the origin of axis, we perform a step of  a fixed length
$R_0$ with an angle $n \pi/(2(L-1))-\pi/8$
if the $n$th bit ($0 \le n \le L$) of genotype $x$ has value one.    
In this way one locates the master sequence (all zeros) 
 at the origin; the strains with one bad gene, distributed
according to the bad gene position
at distance $R_0$; the strains with two bad genes
at an approximate distance 
of $2 R_0$, and so on.
An example of the resulting
hypercube for $L=4$ is shown in Figure~\ref{fig:Hyper}.

We are interested in computing the critical  values of parameters
for the transition between one single 
quasi-species to coexisting quasi-species (speciation).

We obtain from the approximate analysis of Appendix~A
that the crucial parameter in the limit $\mu \rightarrow 0$ is
the quantity $G=(J/R)/(b/r)$, which is the ratio of two
quantities, one related to the strength of inter-species interactions
($J/R$) and the other to intra-species ones ($b/r$).
We observe  ,in good agreement with the analytical approximation see
Eq.~(\ref{Gc}), 
if $G\,m > G_c(r/R)$  
several quasi-species coexist, otherwise only the master sequence
quasi-species survives. Here $m$ is the asymptotic 
average population size that is usually close to one at the
transition point.
The approximate analysis also shows that 
this  transition does not depend on the mutation
rate in the first approximation.
In Figure~\ref{fig:Spot} a distribution with multiple
quasi-species is shown.

We can characterize the speciation transition 
by means of the entropy $S$ of the asymptotic distribution, 
\[ 
    S = - \sum_x p(x) \ln p(x)
\]
which increases in correspondence of the appearance of multiple
quasi-species. 

In Figure~\ref{fig:Entropy} we characterize 
this transition as an increase of the entropy as function of
$G\,m$. We can locate the transition at a value $G\,m \simeq 2.25$, 
analytical approximation predicts $G_c(0.1)\simeq 2.116$. 
The entropy however is quite sensible to
fluctuations of the master sequence quasispecies (which embraces the
largest part of distribution), 
and it was necessary to average over several (15) runs in order
to obtain a clear curve; for larger values of $\mu$ it was impossible
to characterize this transition. A quantity which is much less 
sensitive of fluctuations is the
average square phenotypic distance from the master sequence
 $\overline{g(x)^2}$ 
\[
 	\overline{g(x)^2}=\sum_x g(x)^2 p(x).
\]
In Figure~\ref{fig:x2} (left)  we characterize the speciation transition by
means of $\overline{g(x)^2}$, and indeed a single run was sufficient,
for $\mu=10^{-3}$. For much higher mutation rates ($\mu=5\; 10^{-2}$) the
transition is less clear, as shown in Figure~\ref{fig:x2} (right),
but one can see that the transition point is substantially
independent of $\mu$, as predicted by the approximate theory,
Eq.~(\ref{Gc}).  

Another interesting phenomenon is the meltdown transition, 
for which the mean field theory predicts 
extinction  if
$\overline{\pi}\le 1/2$, see Eq.~(\ref{mequation}). 
In Figure~\ref{fig:melt} we report the result of one simulation in
which the extinction  is induced by the increase of the mutation rate
$\mu$. One can notice that the transition is discontinuous, $m$ jumps
to zero from a value of about 0.15, and that the critical value of
$\overline{\pi}$ is larger that the predicted one.   
This discrepancy can be caused by fluctuations, due to the finiteness
of population. 

\section{Conclusions\label{sec:Conclusions}}

We have studied a microscopic model of a simple ecosystem that exhibit the
mutational meltdown effect and speciation phenomena. The size
of the population is not hold constant; we found that
in the mean filed approximation  this quantity
is ruled by a logistic equation, with the average fitness of
population as control parameter. 
The model includes the competition among individuals, and this
ingredient is considered fundamental for the speciation phenomenon in
a smooth fitness landscape. This transition does not depend on the mutation
rate provided that this rate is small. We are able to
obtain  analytical approximations for the onset of both transitions.  

\nonumsection{Acknowledgements}
We thanks P. Palmerini for having installed MPI on the cluster of linux
machines and for his well written help pages on parallel computing.
We are also grateful to D. Stauffer for his rude incentives and suggestions. 
M. Bezzi thanks the Department of Applied Mathematics for
hospitality. 

\nonumsection{Appendix A: Analytical approximations\label{sec:MeanField}}
Some analytical results can be obtained by considering the dynamics
only in the phenotypic space. Let us consider the case of the
phenotype that depends only on the number of bits (say, good genes)
in the genotype, i.e.\  a highly degenerate phenotypic space. 
One should introduce
the multiplicity factor (binomial) 
of a given phenotype, which can be approximated to a
Gaussian;
however if one works in the neighborhood of 
the most common chemical composition, the
multiplicity factors are nearly constants. Another reason for not
using the multiplicity factor is that we have not yet been able to
derive analogous results with it. 

An instance of an application of a similar (sub-)space in the
modeling of the evolution of real
organisms is  given by a
repeated gene (say a tRNA gene): a fraction of its 
 copies can mutate, linearly varying the
fitness of the individual with the ``chemical composition''
of the gene\protect\cite{BagnoliLio}. This degenerate case has been
widely studied (see for instance Alves and Fontanari 
(1996)\protect\cite{Alves}); 
Another example is given by the level of catalytic
activity of a protein.  A non-degenerate
 space has also been used for modeling the
evolution of RNA viruses on HeLa cultures.\protect\cite{Kessler}

From now on we shall indicate with $x$ both the phenotype and the
genotype, and consider it as an integer number. To maintain a bit of
the original multiplicity, we extend the range of $x$ to negative
values, while keeping the master sequence at $x=0$. 

We compute from Eq.~(\ref{peq})  the values of parameters that allow the
coexistence of different species. We look for a solution $p(x)$ formed by
the sum of delta peaks ($\mu \rightarrow 0$ limit) centered at $y_k$.

\[
p(x)=\sum_k\gamma _k\delta (x-y_k)\equiv \sum_kp_k
\]

The weight of each quasi species is $\gamma_k$, i.e.
\[
	 \int p_k(x) dx = \gamma_k, \qquad\sum_{k=0}^{L-1} \gamma_k = 1.
\]

The evolution equation for  $p_k$ is:

\[
p_k^{\prime }(x)=\frac{\pi(y_k)}{\overline{\pi}}p_k(x)
\]

The stability condition of the asymptotic distribution ($p_k^{\prime }(x)=
p_k(x)$) is 
 either
 $\pi(y_k) = \overline \pi = \text{const}$
(degeneracy of maxima) or $p_k(x)=0$ (all other points). In other terms one
can
say that in a stable environment the fitness of all individuals is the
same,
independently on the species. 

The position $y_k$ and the weight $\gamma_k$ of the quasi-species
are given by $\pi(y_k) = \overline \pi = \text{const}$ and 
$\left.{\partial \pi(x)}/{\partial x}\right|_{y_k} = 0$, or, in terms of the
fitness $H$, by 
\[
	h(y_k) - J m \sum_{j=0}^{L-1} K\left(\dfrac{y_k-y_j}{R}\right)
		 \gamma_j = \text{const}
\]
\[
	h'(y_k)  - \dfrac{J m}{R}\sum_{j=0}^{L-1} K'\left(\dfrac{y_k-y_j}{R}\right)
	 \gamma_j = 0
\]

Let us compute the phase boundary for coexistence of three species for two
kinds of kernels: the exponential  one ($\alpha=1$)
and a Gaussian one ($\alpha=2$). 
Numerical simulations show that the results
are qualitatively independent on the exact form of the static fitness,
providing it is a smooth decreasing function. 

Due to the symmetries of the problem, we have one quasi-species at $x=0$
and
two symmetric quasi-species at $x=\pm y$. Neglecting the mutual influence
of
the two marginal quasi-species, and considering that $h'(0) = K'(0)=0$, 
$K'(y/R) = -K'(-y/r)$, $K(0)=J$ 
and that the three-species threshold is given by $\gamma_0=1$ and
$\gamma_1=0$,
 we have 
\[
	\tilde{b}\left(1-\dfrac{z}{\tilde{r}}\right) 
				- K(z) = -1, \qquad
	\dfrac{\tilde{b}}{\tilde{r}} + K'(z) = 0,
\]
where $z=y/R$, $\tilde{r} = r/R$ and $\tilde{b} = b/J$. 
We introduce the parameter $G=\tilde{r}/\tilde{b}=
(J/R)/(b/r)$, that is the ratio of two
quantities, one related to the strength of inter-species interactions
($J/R$) and the other to intra-species ones ($b/r$). 
In the following we shall drop the tildes for convenience.
Thus
\[
 r - z -m G \exp\left(-\dfrac{z^\alpha}{\alpha}\right) = -m G,
\qquad m G z^{\alpha-1}\exp\left(-\dfrac{z^\alpha}{\alpha}\right) = 1;
\]
Where $m$ can be obtained from Eq.~(\ref{mequation}),
\begin{equation}
m =\frac{2\overline{\pi}-1}{\overline{\pi}^2} \qquad \mbox{with} \qquad 
\overline{\pi}=\pi(0) = \dfrac{e^{\beta h_0}}{1+e^{\beta h_0}},
\end{equation}
and thus 
\[
 m= 1 -  e^{- 2 \beta h_0}.
\]

For $\alpha=1$ the coexistence condition never holds, except 
for  $G\;m=1$ and $r=0$, i.e.\ a
 flat landscape ($b=0$) with infinite range interaction ($R=\infty$). 
Thus we suppose that the speciation transition is not present
also for less steep potentials, such as power laws.  

For $\alpha=2$ the coexistence condition is given by
\[
	z^2 -(m G+r)z + 1 = 0,
\qquad
	m Gz\exp\left(-\dfrac{z^2}{2}\right) = 1.
\]
One can solve numerically this system and obtain the boundary 
$G_c(r)$ for the coexistence. In the limit $r \rightarrow 0$ (static
fitness
almost flat) and $ \beta h_0 \gg 1$ (i.e. $m \simeq 1$) one has 
\begin{equation}
	G_c(r) \simeq G_c(0) - r \label{Gc},
\end{equation}
with $G_c(0) = 2.216\dots$. 
Thus for $G>G_c(r)$ we have coexistence of three or more quasi-species,
while 
for $G<G_c(r)$ only the fittest one survives.
The limit $ \beta h_0 \gg 1$ ($m \simeq 1$) is not a restrictive
condition from a theoretical point of view , in fact we can always stay 
in this approximation modulating $\beta$; but we get $\pi$ almost
constant and equal to one, therefore there is a shortage of empty
cells and the evolution can take much longer times. 

\nonumsection{Appendix B: Monte Carlo Algorithm}
We describe here the essentials of the implementation of the model in
FORTRAN (although we use C language).

The implementation of the model can be done in a direct way, but since the
coupling due to competition is global, the simulation time grows as $N^2$,
where $N$ is the number of individuals present in the environment. 
A way of speeding up a little the simulation is that of performing the
computation of the fitness using Eq.~(\ref{Hn}) instead of
Eq.~(\ref{fitnessH}). 

The cellular automaton space is the vector
{\tt integer env(0:N-1)}. The surviving individuals always occupy the
first {\tt M} positions (starting from $0$): insertions always occurs
at position {$\tt M$} ({\tt M} is thereafter incremented), and
deleted individuals are overwritten by the genotype in position {\tt
M-1} ({\tt M} is thereafter decremented).

We also use three other vectors,  
{\tt integer strain(0:N-1)}, {\tt integer distr(0:L2-1)} and
{\tt real fit(0:L2-1)}, where {\tt L} is the genome
length and {\tt L2=2**L}.
 The first vector contains {\tt NS} entries corresponding to 
 each instance of a different genome in
the environment. This vector is needed to perform sums over all genotypes
without scanning {\tt env}. The vector  {\tt distr}
contains the number of instances of a given genome in the environment, and 
the vector {\tt fit} its survival probability. 

The static fitness $h$ and interaction matrix $\J$ are stored in the
vectors {\tt real h(0:L2-1)} and {\tt real J(0:L2-1,0:L2-1)}, which
are filled at the beginning.  The central loop of the  evolution
algorithm  is the following:

\begin{verbatim}
C
C assume that strain and distr are already OK 
C and compute fit
C
  do i = 0, NS-1
   ig = strain(i)
   fit(ig) = 0
   do j = 0, NS-1
     jg = strain(j)
     fit(ig) = fit(ig) + J(ig,jg)*distr(jg)
   end do
   fit(ig) = fit(ig)/N + h(ig)
   fit(ig) = exp(fit(ig))/(1+exp(fit(ig)))
  end do
C
C  clear strain and distr
C
  do i = 0, NS-1
   ig = strain(i)
   distr(ig) = 0
  end do
  NS = 0
C
C  survival 
C
  i=0
  do while (i .lt. M)
   r = rnd(iseed)
   if (fit(env(i)) .lt. r) then        ! don't survives
    env(i) = env(M-1)
    M = M-1
   else                                ! survives
    if (distr(env(i)) .eq. 0) then     ! first instance of genome
     strain(NS)=env(i)
     NS = NS+1
    end if 
    distr(env(i)) = distr(env(i)) + 1
    i = i+1
   end if
  end do
C
C  reproduction
C
  M1 = M
  do i=M1, N-1
   j = int(rnd(iseed)*N)
   if (j < M1)                         
    if (rnd(iseed) .lt. mu) then      ! reproduction
     env(M) = ieor(env(j),2**(int(rnd(iseed)*L))  ! this is a XOR
    else
     env(M)=env(j)
    end if
    if (distr(env(M)) .eq. 0) then     ! first instance of genome
     strain(NS)=env(M)  
     NS = NS+1
    end if    
    distr(env(M)) = distr(env(M)) + 1
   end if
   M=M+1
  end do
\end{verbatim}     

The program has been implemented on a CRAY T3E and on a cluster of
Linux machines using MPI. Since the code is not well parallelizable,
due to the long range interactions and on the updating scheme, we
have parallelized on the control parameters and on different runs.
In other words we have launched a  copy of the program in parallel on
a different CPU, and the results have been collected using MPI. In
this way also a cluster of machines with relatively slow 
connections (ethernet) can be used  as a supercomputer. 

\nonumsection{References}

\end{document}